# Decentralized Prediction Market without Arbiters

Iddo Bentov[1], Alex Mizrahi[2], and Meni Rosenfeld[3]

[1] Cornell University, `iddobentov@cornell.edu`
[2] chromaway.com, `alex.mizrahi@gmail.com`
[3] Israeli Bitcoin Association, `meni@bitcoin.org.il`

**Abstract.** We consider a prediction market in which all aspects are controlled by market forces, in particular the correct outcomes of events are decided by the market itself rather than by trusted arbiters. This kind of a decentralized prediction market can sustain betting on events whose outcome may remain unresolved for a long or even unlimited time period, and can facilitate trades among participants who are spread across diverse geographical locations, may wish to remain anonymous and/or avoid burdensome identification procedures, and are distrustful of each other. We describe how a cryptocurrency such as Bitcoin can be enhanced to accommodate a truly decentralized prediction market, by employing an innovative variant of the *Colored Coins* concept. We examine the game-theoretic properties of our design, and offer extensions that enable other financial instruments as well as real-time exchange.

## 1 Introduction

A prediction market (PM) enables its participants to continuously place bets on the outcome of uncertain future events. As the PM is transparent and provides price discovery, each participant can take into consideration the current market price for outcomes of events, and attempt to make informed decisions regarding whether to buy or sell shares in such outcomes. Another use of PMs is in hedging positions. An individual may buy a prediction not because she believes that the event will happen, but because it would have a negative effect on her. She thus reduces her risk by betting on the event, anticorrelating with her current position. Further, PMs function as a useful forecasting tool even for non-participants, because predictions that are made when traders risk their own money have proven to be more accurate than polls and other methods [2].

The decentralized structure of the Bitcoin [17] network implies that its soundness does not require reliance on trusted parties, and that its participants can operate anonymously [13–16] if they take appropriate precautions. By utilizing *Colored Coins* [20] protocols, a decentralized stock exchange and other financial services can be integrated with Bitcoin. Similarly, "meta"-protocols such as the Counterparty [11] and Omni [18] layers[4] can provide more advanced financial services. Thus, one may regard it to be of interest to explore whether a decentralized PM can also be deployed on top of Bitcoin.

At first glance, it may seem that a decentralized exchange of assets poses less of a challenge than a decentralized PM. This is because the relevant aspects when trading an asset are just whether the issuer of the specific asset is reputable, and whether the trading platform is secure. Some assets may not require any reliance on reputation, e.g. an asset that gives ownership rights over a digital art item (including the right to present it at a gallery), which can then be traded in an atomic fashion. By contrast, even though a PM only deals with digital information, a fully decentralized PM requires a broad agreement regarding the outcomes of events.

---
[4] Each of them reached a market cap greater than $20 million in 2014, see `http://coinmarketcap.com/`.

Indeed, the work of [7] constructs a PM via a cryptocurrency of the Bitcoin mold, but it relies on trusted arbiters to decide the correct outcome of events. An alternative presented in [7] is to let the "miners" who perform the *Proof of Work* computations register their votes on the outcomes of events in the blocks that they solve. However, this alternative raises significant concerns, which stem from the fact that miners in a decentralized cryptocurrency can operate anonymously. Consider, for example, an obscure event that is relevant only to a small village. Some faction of this village can try to bribe miners to vote for their preferred outcome. Ideally, miners would be disincentivized from voting incorrectly as it entails the risk of losing the block reward in case their solved block is rejected by honest miners. For this to happen, honest miners would need to parse every obscure event description and keep up with the real-world outcomes of such events, which is impractical. Hence, choosing a trusted entity as arbiter in accord with hierarchical certification (cf. [7, Section 5.3]) is probably a better option for a PM of this kind.

One may ask why it is of value to decentralize all the aspects of a PM. Some of the possible reasons are as follows:

– For arbiters, credibility is inversely correlated with susceptibility. An anonymous arbiter is probably untrustworthy, while a well-known arbiter can be pressured by hostile elements to not resolve an event correctly.
– Eliminating the need for arbiters makes it easier to bet on events that extend over a very long time period (e.g., "Texas will secede from the U.S. before the year 2030"), or even events with unbounded time (e.g., "The State of Jefferson will be created out of California and Oregon before Texas secedes from the U.S."). It is desirable to let the market assign probabilities to such outcomes in a continuous fashion while relevant occurrences in the world unfold, without running the risk that a designated arbiter will not be alive or no longer be (the *only* one) in possession of her secret signing key at the time when the outcome is resolved.
– Anonymous traders may make predictions on interesting events that a traditional PM does not tend to accommodate. For example, "Street gang #1 will win the turf war in which they swore to expel street gang #2 from region x before the year 2018". Market participants might not be able to agree in advance on a trustworthy arbiter for this event, even though the outcome can be agreed upon by impartial observers and hence suitable to be decided by a decentralized PM.
– Reputable arbiters may expect to be compensated for the service that they provide, in part because they need to take precautions to secure their secret signing key. This implies that market participants will need to pay fees that go to the trusted arbiters, on top of the fees that are paid to the miners.

- In case the designated arbiter makes the wrong call for an event resolution, her decision becomes irreversible according to the protocol rules of a semi-decentralized PM [7]. Thus, shares of the winning outcome that are still in circulation are unfortunately worthless. This stands in contrast to a fully decentralized PM, in which market forces will re-adjust the value of the shares as the mistaken outcome becomes known.[5]

These reasons add to the obvious observation that designated arbiters may be malicious or willing to be bribed. For instance, a corrupt arbiter may stonewall and refuse to sign the correct resolution of an event until she receives extra money on the side. The corrupt arbiter may also stock up on cheap shares of an unlikely outcome, then rule in favor of that outcome and in effect steal money from other traders.

In Section 3 we discuss the conditions under which our fully decentralized PM scheme is likely to work well, and conditions under which a PM with trusted arbiters may be more appropriate.

### 1.1 Prediction markets with anonymous participants

An anonymous marketplace with or without trusted arbiters can facilitate insider trading and other kinds of fraud that are less probable in non-anonymous setting.

E.g., a goalkeeper can secretly buy shares that predict that her team will lose a soccer game, then concede goals on purpose and profit. Still, even in a non-anonymous PM the goalkeeper may ask someone else to buy the shares and later divvy up the profits between them, hence the issue boils down to the observation that an anonymous marketplace allows fraudsters to operate with less friction.

Therefore, it is safer to bet on events whose significance is likely greater than their trade volume, particularly in the case of a PM with anonymous participants.

See for example [9] for further discussion and analysis of outside incentives.

### 1.2 Related work

The work of [7] presents a cryptocurrency protocol for a PM that is decentralized in the sense that anyone can inject liquidity for betting on new or existing events, but centralized in the sense that it depends on trusted arbiters to decide the outcomes of events. Moreover, [7] presents a decentralized matching platform for PM trading directly on the cryptocurrency network. In Appendix A we outline how it is also possible to construct a trading platform that is suitable for real-time trades.

The Truthcoin [22] and Augur [19] projects attempt to build a different variant of a decentralized PM, where holders of tradeable "reputation" cryptocurrency take over the role of trusted arbiters in deciding outcomes of events. This is done via quite intricate voting methods in which all holders of these reputation coins may cast their votes for each event resolution, voters who agree with the majority earn fees, and voters who end up in the minority may suffer a loss. One aspect that neither *Truthcoin* nor *Augur* try to decentralize is the initial issuance of reputation coins by means of an auction or an IPO (cf. [6, Section 4] and [8, Section V.B]). By contrast, in our PM protocol the outcomes of events are decided by market forces rather than by votes, hence there is no need for an IPO that would potentially enrich the parties that initiate the PM system.

---

[5] An example of a mistaken ruling is the 2012 Iowa caucus incident at https://en.wikipedia.org/wiki/Intrade#Disputes.

## 2  Mechanism

The *Colored Coins* concept [20] allows Bitcoin to support non-fungible assets rather than only fungible coins. This means that "tagged" or "colored" coins can be sent and received on the Bitcoin network. Thus, if Alice has a portfolio of $\{(5, \texttt{red}), (6, \texttt{blue})\}$ coins, she can send $(1.9, \texttt{red})$ coins to Bob's address and have $\{(3.1, \texttt{red}), (6, \texttt{blue})\}$ coins remaining.

The PM system that we hereby construct is based on Bitcoin, with all assets colored according to the fixed form $(\texttt{amount}, \texttt{bet}, \texttt{history})$. Initially, the system has uncolored assets $(\texttt{amount}, \bot, \emptyset)$, that can be used in exactly the same way as ordinary bitcoins. For example, if Bob has $(9, \bot, \emptyset)$ coins, he can send $(1.2, \bot, \emptyset)$ coins to Alice's address and have $(7.8, \bot, \emptyset)$ coins left.

To allow everyone to participate in the PM in a fully decentralized fashion, we define three types of special transactions, as follows.

**Creating a prediction pair.**  Anyone can execute a special *outcome-split* transaction that transforms her $(\texttt{amount}, \bot, \texttt{history})$ asset to

$$\{(\texttt{amount}, \texttt{Yes:eid}, \texttt{history}), (\texttt{amount}, \texttt{No:eid}, \texttt{history})\},$$

where $\texttt{eid}$ is some particular event-id that is derived via

$$\texttt{eid} = \texttt{hash}(\text{"Textual description of an event"}).$$

We assume that $\texttt{hash}()$ is a cryptographic hash function. These split Yes/No shares can now be transferred as is the case with colored coins. E.g., Alice may split $(m, \bot, \emptyset)$ using event-id $\texttt{eid}_0$, then send $(2/3 \cdot m, \texttt{Yes:eid}_0, \emptyset)$ shares to Bob, and remain with $\{(1/3 \cdot m, \texttt{Yes:eid}_0, \emptyset), (m, \texttt{No:eid}_0, \emptyset)\}$ shares in her possession.

**Redeeming a prediction pair.**  Anyone in possession of $(\texttt{amount}, \texttt{Yes:eid}, h_1)$ shares and $(\texttt{amount}, \texttt{No:eid}, h_2)$ shares is allowed to execute a special *outcome-combine* transaction that transforms these shares to $(\texttt{amount}, \bot, h_1 \cup h_2)$.

Hence, no matter what are the current market value of $(\texttt{amount}, \texttt{Yes:eid}, \emptyset)$ and $(\texttt{amount}, \texttt{No:eid}, \emptyset)$ separately, their combination is always worth $(\texttt{amount}, \bot, \emptyset)$ ordinary coins.

**Forcing an encumbered history.**  Anyone can execute a special *outcome-force* transaction that transforms her $(\texttt{amount}, \texttt{Yes:eid}, \texttt{history})$ asset to $(\texttt{amount}, \bot, \texttt{history} \cup \{\texttt{Yes:eid}\})$.

Likewise, anyone can transform her $(\texttt{amount}, \texttt{No:eid}, \texttt{history})$ asset to $(\texttt{amount}, \bot, \texttt{history} \cup \{\texttt{No:eid}\})$.

Let us elaborate on these mechanisms by providing several examples. See the accompanying Figure 1 for an illustration.

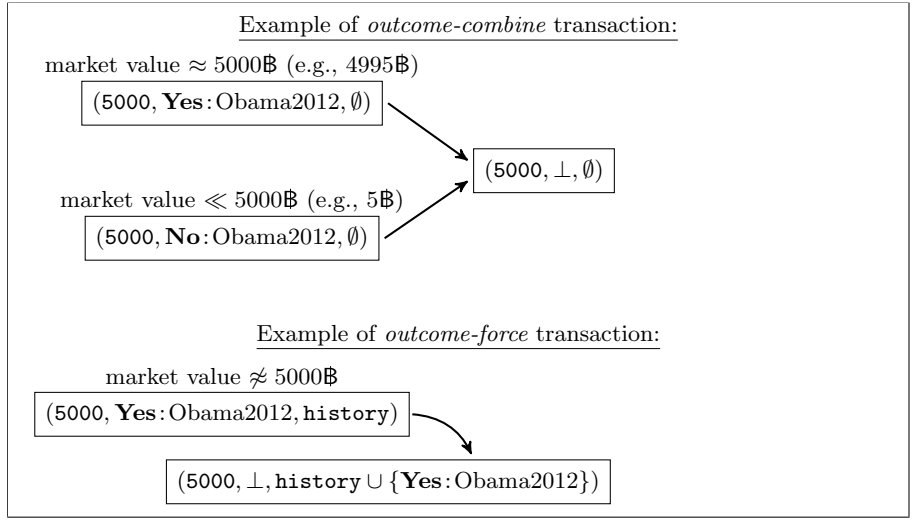

**Fig. 1.** Special transactions for event resolution.

**Exemplary scenario 1.** During 2011, Alice believes that President Obama will win the 2012 presidential election. She computes Obama2012=hash("Barack Obama will win re-election in 2012") and executes *output-split* to transform $(5000, \bot, \emptyset)$ ordinary coins that she possesses to

$$\{(5000, \texttt{Yes:Obama2012}, \emptyset), (5000, \texttt{No:Obama2012}, \emptyset)\}.$$

Suppose that the market believes that President Obama has 70% probability to win re-election. Alice trades her $(5000, \texttt{No:Obama2012}, \emptyset)$ shares for $(1500, \bot, \emptyset)$ ordinary coins, since $30/100 \cdot 5000 = 1500$. After President Obama wins re-election, the market price of No:Obama2012 plummets to 0.001 coins per share. Hence, Alice buys $(5000, \texttt{No:Obama2012}, \emptyset)$ shares for 5 coins, and then uses the $(5000, \texttt{Yes:Obama2012}, \emptyset)$ shares that she kept to execute *outcome-combine* and earn 5000 coins back. Alice's total gain is $1500 - 5 = 1495$ coins.

**Exemplary scenario 2.** During 2011, Alice wishes to risk her entire wealth of 5000 coins by betting in favor of President Obama's re-election. The market assigns 70% probability to this event. Alice trades her $(5000, \bot, \emptyset)$ coins for $(7142.8, \texttt{Yes:Obama2012}, \emptyset)$ shares on the market, since $7142.8 \cdot 70/100 \approx 5000$. After President Obama wins re-election, the market price of Yes:Obama2012 rises to 0.999 coins per share, hence Alice sells her $(7142.8, \texttt{Yes:Obama2012}, \emptyset)$ shares for $(7135.7, \bot, \emptyset)$ coins. Alice's total gain is $7135.7 - 5000 = 2135.7$ coins.

The difference between scenarios 1 and 2 demonstrates that traders who provide the initial liquidity to the market need to commit more funds than the traders who join later, thus it can be reasonable for early traders to expect a small premium over the market price. This premium can be materialized in the form of a slightly wider bid-ask spread.

**Exemplary scenario 3.** During 2011, Alice wishes to bet in favor of President Obama's re-election, and executes *output-split* to transform her $(1000, \bot, \emptyset)$ coins to $(1000, \texttt{Yes:Obama2012}, \emptyset)$ shares and $(1000, \texttt{No:Obama2012}, \emptyset)$ shares. The market believes that the price of a $\texttt{No:Obama2012}$ share is $^{30}/_{100}$ coins. Hence, Alice sells her $(1000, \texttt{No:Obama2012}, \emptyset)$ shares for $(300, \bot, \emptyset)$ coins. After President Obama wins re-election, Alice wishes to buy $(1000, \texttt{No:Obama2012}, \emptyset)$ shares in order to execute *outcome-combine*, but traders who hold these shares demand an unreasonable high price of $^{20}/_{100}$ coins per $\texttt{No:Obama2012}$ share. Alice declines to pay such an excessive price, and instead executes an *outcome-force* transaction to transform her $(1000, \texttt{Yes:Obama2012}, \emptyset)$ shares to $(1000, \bot, \{\texttt{Yes:Obama2012}\})$ coins with encumbered history. Thus, Alice presumes that since all reasonable people should agree that President Obama won in 2012, she will be able to pay with these encumbered coins at the grocery store, etc. For instance, a store may accept Alice's payment of $(803, \bot, \{\texttt{Yes:Obama2012}\})$ for an item that normally costs 800 coins.

As scenario 3 alludes to, the intuitive reason for supporting an *outcome-force* operation is that it serves as a deterrent against traders who would demand an excessive price for their losing shares, by offering an alternative that removes the dependence on such misbehaving traders. The game-theoretic implications of *outcome-force* are examined with more details in Section 3. Let us note that misbehaving traders can pose problems even with a trusted arbiter who may not decide the outcome until a future date, which implies that unless the traders who hold the losing shares act reasonably, the winning shares would be neither interest-baring nor spendable for a possibly long time period (cf. [7, Section 4.2]).

It is likely that traders will prefer to buy the losing shares for cheap and execute *outcome-combine* to obtain coins with a clean history, rather than execute *outcome-force* and encumber the history of their coins, because nobody wants to run the risk of having unrecognized coins that get declined when they attempt to make payments. Still, some users of the currency might wish to resort to reputable oracles to fetch and thereby recognize widely agreed upon versions of encumbered history. This can be helped via protocol support for hashing event-ids into a single set-id according to a canonical order, so that a set-id can be re-hashed into a larger set when its preimage (that consists of event-ids) is given.

Therefore, when the PM functions properly (as in scenarios 1 and 2), the price of an $(1, \texttt{Yes:eid}, \emptyset)$ share can be interpreted as the probability that the market assigns to the event, since the cost of a losing share will be close to 0.

Another question is why it is needed to execute *outcome-force* instead of simply keeping the winning shares and using them as currency. The reason is that such shares would have to be transacted separately rather than joined with the ordinary coins that the user holds, which entails extra complexity and is not scalable. Also, such shares cannot be used to place bets on a new event, unless they first get converted to a usable format via either *outcome-force* or *outcome-combine*.

Finally, let us note that this PM system relies on a softfork (or hardfork if desired) of the Bitcoin protocol, due to two reasons. First, when colored coins are implemented as an optional layer on top of Bitcoin, miners are oblivious to it, and hence there may not be widespread agreement regarding the coloring rules. Additionally, in optional colored coins layers it is typically the case that one can always "uncolor" a colored coin, which implies that colored coins that exist in the Bitcoin system are always worth at least as much as their uncolored amount. In any case, a protocol fork is needed for a more efficient tagging-based colored coins support (cf. [5,20]), and our reference colored coins implementation with split/combine/encumber operations demonstrates that the overhead for a decentralized PM is minor [10].

## 3 Analysis

We are interested in analyzing what will be the prediction share price when each type of share is traded in the open market. We assume that a pair of "+" share and "–" share can always be exchanged for 1 BTC. In this abstract model, we further assume that due to agreement about which prediction was correct, a "+" share will be worth $p$ BTC even if it is never combined with a "–" share, while a "–" share will be worthless without the possibility to combine it.

The parameter $p$ can be regarded as the probability that the Bitcoin miners and full nodes will form a new consensus rule (by means of a softfork) that cleanses the encumbered coins corresponding to the prediction, thus transforming them into unencumbered coins. For example, the majority of miners will probably agree that it is reasonable to cleanse the aforementioned `Obama2012` event if many such encumbered coins are in circulation. However, in case the miners are unreasonable an wish to regard another candidate as the winner of the 2012 elections (contrary to what the rest of the population thinks), our analysis will unfortunately reflect that by assigning the higher value to what ought to have been "–" shares. Notice that if the event description has some ambiguity, then even reasonable actors may fail to reach consensus. For example, an unambiguous event description for the U.S. presidential election in 2000 could have been "Al Gore will be inaugurated as the 43rd President of the United States".

Let us note that there exists a significant difference between letting miners have the power to vote on outcomes of all events in the blocks that they solve, and the above possibility of miners reaching consensus to cleanse the shares of an old event that are still in circulation. The problematic nature of the first method is discussed in Section 1. By contrast, the second method is a deliberate process that can be done in phases where in an initial phase the miners express willingness to support the supposedly benign fork, and in a latter phase the fork becomes operational. This method of deliberation to reach consensus has already been deployed in Bitcoin several times, in particular for the benevolent P2SH [1] and CLTV [23] forks. Therefore, in the case of well-known events for which there is wide agreement on the outcome among the general population, the decision to cleanse "+" shares can be a suitable candidate for a protocol fork.

It can also be appropriate to regard the probability $p$ as corresponding to other conditions that are easier to meet, for example that a quorum of reputable oracles (that payment processors can utilize as in Section 2) consider "+" shares to be indistinguishable from unencumbered coins. The downside of such a condition is that it relies on a system with some centralized elements, rather than a fully decentralized system.

Other possibilities include deciding the outcome via an algorithm that was not yet known at the time when the prediction was made, or via measurements that rely on physical data and thus cannot be scripted in the cryptocurrency.

Let us stress that the most basic condition is that a user simply consults with herself before accepting an encumbered coin as payment, since popular event descriptions (e.g., "Barack Obama will win re-election in 2012") can be easy enough to consider. Therefore, $p > 0$ should hold even without reliance on extra mechanisms such a miners' fork or reputable oracles, though such mechanisms can help in making $p$ larger.

Generally, the parameter $p$ can thus be considered to be the expected price of a "+" share, where the expectation is taken over all the events that can influence the worth of the "+" shares.

Hence, this is essentially a situation known in game theory as the "glove game" [3]. A common method of analyzing cooperative games like this is the Shapley value [3,21], which essentially gives a stable evaluation of each participant's assets. A coalition of $k$ players with a "+" share and $\ell$ players with a "–" share has a total value of $pk + (1-p)\min(k,\ell)$; so if there are $m$ "+" players and $n$ "–" players, the Shapley value of a "–" player is given by:

$$v_- = \frac{1-p}{(m+n)!} \sum_{i=1}^{m+n} \sum_{j=0}^{\lfloor i/2 \rfloor - 1} (m+n-i)!(i-1)! \binom{n-1}{j}\binom{m}{i-j-1}.$$

And the shapley value of a "+" player is

$$v_+ = \frac{mp - nv_- + (1-p)\min(m,n)}{m}.$$

For example, if $p = 1/10$, $m = 30$ and $n = 25$, then $v_- = 0.670012$ and $v_+ = 0.329988$. As we see in this case, "–" shares actually have the higher value, because the oversupply of "+" shares implies that the holders of those shares have less bargaining power, and $p = 1/10$ is too small to compensate for that. By contrast, $p = 3/4, m = 30, n = 25$ result in $v_- = 0.186114$ and $v_+ = 0.813886$.

There is an economics phenomenon of destroying assets (often food) in order to increase the price of the stock that was kept. While counterintuitive, there are market conditions in which this can actually increase the overall profit. It is interesting to consider whether a similar phenomenon can happen here. Let us note that with Bitcoin and similar cryptocurrencies, players can indeed destroy assets that they control in a publicly verifiable way, by sending an unspent output to a script that always returns `False`.

In normal circumstances this should not happen. If a player chooses to "burn" some of her coins, this will increase the Shapley value of her remaining coins – but not so much that her total value will increase. This is because the Shapley value, in a way, considers all possible negotiation tactics of the different players, and if there was a way to gain from burning coins, it should already be accounted for in the original Shapley value.

But this can happen in the case of incomplete information and erroneous assumptions by the players. For example, assume there are 100 "+" shares and 100 "–" shares, with $p = 0$. Most players assume that there are 200 players with 1 share each; they base their trading activity on this assumption, and this results in a market value equal to the Shapley value for this game. However, unbeknownst to them, there is actually a single player in possession of all 100 "+" shares. If she decides to visibly burn some shares and keep only $m$, and the market reacts naively by calculating the Shapley value for a new game with a reduced number of players, her total value as a function of the coins she keeps is given in Figure 2.

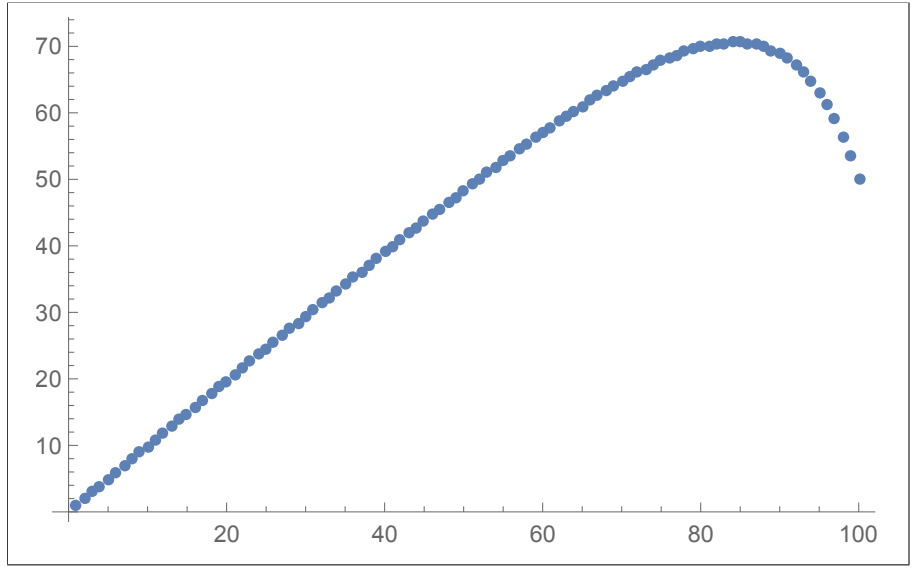

**Fig. 2.** Total value after the destruction of shares.

We can see that as this player starts burning shares, the rise in the value of each is steep enough to increase her total value. She will get the optimal value of 70.5882 if she keeps only 84 shares. Thus, in case this player had less than 84 shares to begin with, it would be disadvantageous for her to burn shares.

Notice that the benefit in burning "+" shares depends on the parameter $p$. This can be seen by noting that the value of each "+" share is given by $v_+ = p \cdot 1 + (1-p) \cdot s_+$, where $s_+$ is the Shapley value for a "+" share in a game with $p = 0$. To see that the equality holds, consider the event

$E_+ = \{$a player with "+" completes a pair in a random permutation of the players$\}$.

According to the definition of the Shapley value (see also Figure 3), we have

$$s_+ = \Pr(E_+)$$
$$v_+ = \Pr(E_+) \cdot 1 + (1 - \Pr(E_+)) \cdot p = p \cdot 1 + (1-p) \cdot s_+.$$

Therefore, destroying all but $x$ shares implies a revenue of $x \cdot (p \cdot 1 + (1-p) \cdot s_+(x))$, i.e., with $s_+$ as a function of $x$. We can thus see that as $p$ tends towards 1, the destruction of "+" shares becomes counterproductive.

| "+" | "+" | "−" | "+" | "−" | "+" | "+" | "−" | "+" | "+" |
|-----|-----|-----|-----|-----|-----|-----|-----|-----|-----|
| 0   | 0   | 1   | 1   | 2   | $p$ | $2p$ | $1+p$ | $1+2p$ | $1+3p$ |

**Fig. 3.** Incremental value of coalitions in glove games with $p = 0$ and $p > 0$.

On the other hand, the decision to burn "−" shares is unaffected by $p$. The reason for this is that each "−" share is worth $v_- = p \cdot 0 + (1-p) \cdot s_-$, with $s_-$ being the Shapley value for "−" in a game with $p = 0$. The rationale for this equality is the same as in the case of $v_+$ above. As we can thus see, burning all but $x$ of the "−" shares implies a revenue of $x \cdot (1-p) \cdot s_-(x)$, and the maximum of this expression does not depend on $p$.

Figure 4 demonstrates the total value that a player with $m$ shares can obtain by not revealing that she possesses the entire supply of the "+" shares. Thus, as in the previous example we assume that there are $m$ individual players who possess one "−" share each, and a single player with all of the $m$ "+" shares. In this figure,

$m = 100$ corresponds to $m(p + s_+(m)(1-p)) = 100(p + \frac{1}{2}(1-p) = 50(1+p)$,

$p = 0$ corresponds to Figure 2,

$p = 1$ corresponds to $m(p + s_+(m)(1-p)) = m$,

and all other values in the range $p \in [0,1], m \in [0,100]$ are plotted.

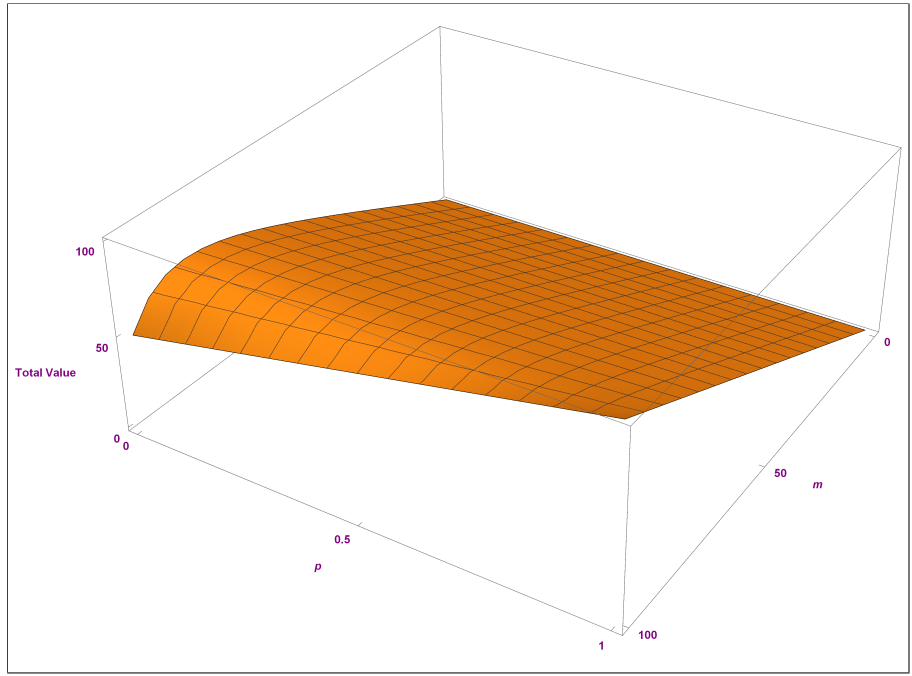

**Fig. 4.** Total value as a function of $p$ and the $m$.

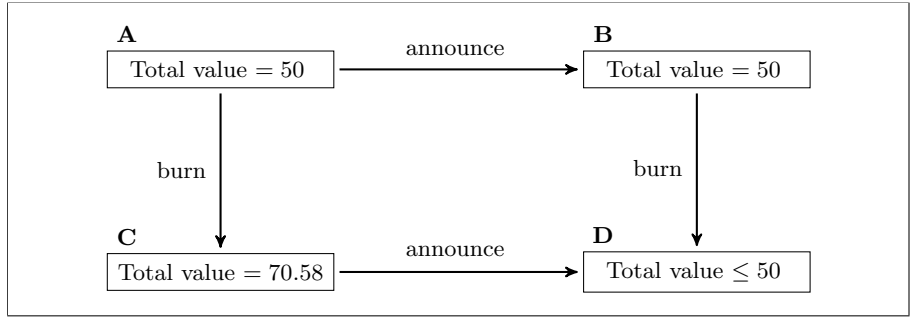

**Fig. 5.** Announcement of ownership vis-a-vis burning of shares.

One may ask whether burning fungible coins (e.g., ordinary bitcoins) could also be profitable for an individual player. Since the value function in this case is different than in the the glove game, the answer is always no. To see this, assume that there are $n$ coins in total that are worth $C/n$ each. Consider a player who possesses $m$ coins, hence her total value is $m \cdot C/n$. In case this player burns $x$ coins, each coin would now be worth (at most) $\frac{C}{n-x}$, and the total value of this player would be $\frac{m-x}{n-x} \cdot C$. Since $m < n \Rightarrow \frac{m-x}{n-x} < \frac{m}{n}$, it follows that burning ordinary coins is always unprofitable.

It is also appropriate to consider whether a player who possesses a large amount of "+" shares can gain an advantage by simply announcing that she controls this entire amount, instead of burning some of her shares. From the technical aspect, such an announcement can be done in Bitcoin and other cryptocurrencies: the player who possesses this amount of "+" shares can publish one common message that is signed with all of the secret keys that control these "+" shares, and thereby prove that these shares have a common owner.

However, due to the fact that the Shapley value takes into account all the possible strategies of the players, such an announcement would in fact have a detrimental effect from the point of view of this player.

To demonstrate, let us consider the same setting that Figure 2 describes. Thus, we assume that $p = 0$, that a single player named Alice has all of the 100 "+" shares, and that the 100 "−" shares are held by 100 individual players. Alice will earn the value of $s_+(100) \cdot 100 = 50$ if she trades her shares on the open market. For a comparison between announcement of ownership and burning of shares, we first note that the diagram in Figure 5 is commutative, i.e., the state **D** can be reached from the initial state **A** by traversing through either state **B** or state **C**. That is to say, Alice can first burn some amount of shares and then announce ownership over the remaining shares, or announce ownership over all of her 100 shares and then burn that same amount of shares, and both cases will result in the same state. This holds because Alice and the other 100 players will have the exact same information after Alice carries out these two actions, hence the resulting Shapley value of "+" and "−" shares will be the same. Next, note that the transition from state **A** to state **B** does not affect Alice's Shapley value, because the symmetric glove game with a single player who has all the "+" shares also gives the values $s_+ = s_- = 1/2$. This can be seen for example by viewing the game as a combination of 100 games where Alice plays against only one player with a single "–" share in each of these 100 games. In these games, Alice's Shapley value is $1/2$, and due to linearity of the Shapley value, it follows that in the original game Alice's Shapley value is $100 \cdot 1/2 = 50$. In state **A**, Alice's Shapley value is also 50, which follows from the definition of the Shapley value and symmetry. On the other hand, as we have seen in Figure 2, by first burning shares (i.e., traversing from state **A** to state **C**), Alice can increase her Shapley value to an optimum of 70.58. Moreover, we note that the edge **B** → **D** only decreases Alice's Shapley value. This is because the strategies that the Shapley value takes into account already include the action of burning shares: if Alice does not settle for the Shapley value and instead defects by burning shares or performing any other action, then her resulting Shapley value will only decrease. In summary, the announcement of the **A** → **B** edge is ineffective, and the announcement of the **C** → **D** edge is detrimental.

Hence, this reasoning serves as an indication that in our anonymous and decentralized PM setting, burning shares (by a player with a large enough amount) is the only action that can potentially be preferable to bargaining in accordance with the Shapley value.

Let us note that there are also other concepts in analysing cooperative games with side payments, such as the core [12]. However, the Shapley value represents an evaluation that is reached from repeated bargaining among the players, as is indeed the case in an open marketplace, and hence its use is reasonable to in the context of our analysis.

It is desired that the decentralized PM will operate in a way that is advantageous towards players who made the correct prediction. Given the above analysis, this becomes more probable in accord with either of the following properties:

1. The parameter $p$ is larger (the closer that $p$ is to 1, the better).
2. The distribution of players who hold "–" shares is more decentralized.

The second property refers to a condition in which there are many players with a relatively small portion of all the "–" shares, while no single player holds a relatively very large portion of the "–" shares.

We can thus conclude that betting on popular events such as presidential elections is more likely to work well in a fully decentralized PM, in comparison to betting on obscure events. It may be preferable to use a semi-decentralized PM with trusted arbiters when betting on events with less popularity, though Section 1.1 should then be taken into account.

## 4 Extensions

We present here add-ons that complement the core PM mechanism of Section 2.

### 4.1 Continuous outcomes

An event description can specify a non-discrete outcome, for instance $e_1 =$ `hash`("The percentage of votes in favor of staying in the European Union in the referendum in country x on January 1, 2018"). After say 45% voted in favor in this referendum, the `Yes:`$e_1$ and `No:`$e_1$ shares should have a market price of $\frac{45}{100}$ and $\frac{55}{100}$ coins per share, respectively.

However, if one opts to encumber e.g. 10 shares of `Yes:`$e_1$ to pay for groceries, then the merchant would need to recognize that this 10 amount is worth 4.5 unencumbered coins, which requires payment processors of higher complexity.

### 4.2 Non-binary outcomes

Section 4.1 can be generalized to a non-binary fixed amount of outcomes, by extending the protocol to support an *outcome-split*($N$) transaction that utilizes the extra parameter $N$ to transforms (`amount`, $\bot$, `history`) to

$$\{(\texttt{amount}, \texttt{1:eid}, \texttt{history}), \ldots, (\texttt{amount}, \texttt{N:eid}, \texttt{history})\}.$$

For instance, Alice can compute $e_2 =$ `hash`("Percentages for top 24 contestants in American Idol season 99: 1=band, 2=girl, 3=boy, 4=other"), and invoke *outcome-split*(4) to transform her $(60, \bot, \emptyset)$ coins to

$$\{(60, \texttt{1:}e_2, \emptyset), (60, \texttt{2:}e_2, \emptyset), (60, \texttt{3:}e_2, \emptyset), (60, \texttt{4:}e_2, \emptyset)\}.$$

Suppose that the market believes that the top 24 will be divided equally between bands, girls, and boys, and Alice believes that the percentage of bands will be much greater than 33%. Alice sells on the market $(60, \texttt{2:}e_2, \emptyset)$ and $(60, \texttt{3:}e_2, \emptyset)$ for $1/3 \cdot 60 = 20$ coins each. If it later turns out that 50% in the top 24 were bands, 25% were girls, and 25% were boys, then Alice buys 60 shares of $\texttt{2:}e_2$ and $\texttt{3:}e_2$ on the market for $1/4 \cdot 60 = 15$ coins each, and executes *outcome-combine*(4) together with the $(60.\texttt{1:}e_2, \emptyset)$ and $(60, \texttt{4:}e_2, \emptyset)$ shares that she kept. Alice's profit is $2(20 - 15) = 10$ coins.

Suppose instead that no boy has reached the top 24, but holders of $\texttt{3:}e_2$ shares demand a price significantly greater than 0 for their supposedly worthless assets. Alice thus buys $(60, \texttt{2:}e_2, \emptyset)$ shares, and executes *outcome-force* to transform $\{(60, \texttt{1:}e_2, \emptyset), (60, \texttt{2:}e_2, \emptyset), (60, \texttt{4:}e_2, \emptyset)\}$ to the encumbered coins $(60, \bot, \{(\texttt{1:}e_2, \texttt{2:}e_2, \texttt{4:}e_2)\})$, which can be regarded to have the same meaning as in Section 2.

### 4.3 Capped contracts for difference

A contract for difference (CFD) is used for betting on the future value of an asset. In decentralized setting, if a certain stock is currently valued at say $200, Alice places a bet that its value in one year will be $290, and the rest of the market places bets that predict (on average) that its value in one year will be $210, then Alice should profit in case the stock's value in a year will be greater than $250, as $210 + \frac{290-210}{2} = 250$. This can be thought of as a generalization of Section 4.2 in which traders place bets on multiple outcomes $\{\ldots, 199, 200, 201, \ldots\}$, but it is infeasible to use the mechanisms of Section 4.2 because the range of possible outcomes is continuous and large.

When we consider some CFD of an asset $x$ where $x$ is traded for example on NYSE, it may make sense to employ the services of NYSE as a trusted arbiter. However, the centralized nature of this approach carries the same implications as described in Section 1. Consider, for example, a CFD for the BTC/USD exchange rate according to one or several predefined Bitcoin exchanges. These exchanges may collapse, or their secret signing keys may leak due to carelessness or malice, etc. By contrast, a decentralized PM can accommodate a CFD for the fair market price of BTC/USD in a way that is resilient to such potential hazards.

The basic prediction mechanism of Section 2 is already enough to support a simple capped CFD variant. To demonstrate this, let us use the following event-id for a capped CFD of an asset $x$ whose price on January 1, 2016 is \$30: $\texttt{e}_3=\texttt{hash}$("Starting from January 1, 2016, the price of asset $x$ will reach \$40 before reaching \$20").

As in the Black-Scholes model [4], by assuming as an approximation that market movements are caused by a large number of traders which are independent and indistinguishable from random, we have that this CFD instrument behaves locally like Brownian motion and thus its price is linear. That is, the price that market participants would assign to $\texttt{Yes:e}_3$ shares is $c/20 - 1$, and the price assigned to $\texttt{No:e}_3$ shares is $2 - c/20$, where $20 < c < 40$ is the current price of the asset $x$. See Figure 6 for an illustration.

Let us note that it is possible to define capped variants of other financial instruments in a similar fashion, e.g., put and call options. In decentralized setting, all such instruments are inherently capped because one cannot earn more than the coins that were used to create an asset (see also Section 4.4). By contrast, standard CFDs and put/call options are uncapped.

A significant drawback of capped CFDs of this form is that holders of shares corresponding to event-id $\texttt{e}_3$ cannot use shares of say $\texttt{e}_4=\texttt{hash}$("Starting from January 1, 2016, the price of asset $x$ will reach \$50 before reaching \$10") for *outcome-combine* operations, which implies that such CFDs will probably not enjoy a market with high liquidity.

### 4.4 Vector CFDs

We now define and explore *vector* CFDs, which can potentially increase the available market liquidity.

Vector CFDs utilize colored coins of the form $(\texttt{amount}, \texttt{eid}, \texttt{V}, \texttt{J}, \texttt{history})$, such that $\texttt{V} = (b_1, w_1, b_2, w_2, \ldots, b_k, w_k)$ with $\sum_{i=1}^{k} w_i = 1$, and $\texttt{J} \in \{1, 2, \ldots, k\}$. The $\texttt{eid}$, $\texttt{V}$, $\texttt{J}$, fields generalize the $\texttt{i:eid}$ field of Section 4.2. The event-id should conform with a format of the type $\texttt{eid} = \texttt{hash}(\text{baseline asset } x)$, where $x$ specifies the identity of an asset such that the market participants wish to place bets on the future price of $x$.

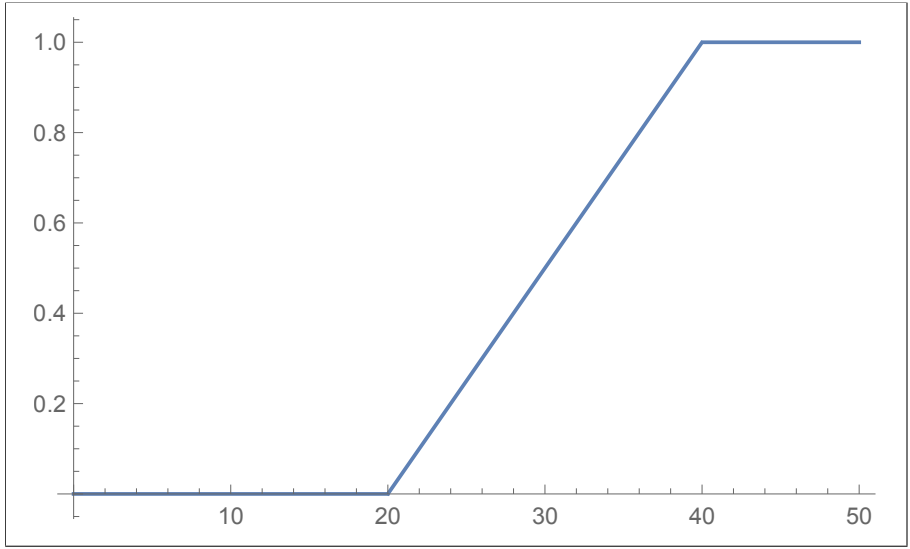

**Fig. 6.** Capped CFD price function.

Variants of the special transactions of the PM system of Section 2 are also used for vector CFDs, as follows.

**Injecting liquidity.** The special transaction *outcome-split* allows $m$ ordinary coins $(m, \bot, \bot, \bot, h)$ to be transformed into $k$ assets $\{z_j = (m, \texttt{eid}, v, j, h)\}_{j=1}^{k}$, where $v = (b_1, w_1, b_2, w_2, \ldots, b_k, w_k)$ and the constraint $\sum_{i=1}^{k} w_i = 1$ holds.

**Soaking liquidity.** The special transaction *outcome-combine* allows the assets $\{(m, \texttt{eid}, v_i = (b_{i,1}, w_{i,1}, b_{i,2}, w_{i,2}, \ldots, b_{i,k_i}, w_{i,k_i}), j_i, h_i)\}_{i=1}^{t}$ to be transformed to ordinary coins $(m, \bot, \bot, \bot, \cup_{i=1}^{t}\{h_i\})$, if the constraints $s_1 = s_2 = \cdots = s_t$ and $\sum_{i=1}^{t} w_{i,j_i} b_{i,j_i} = s_1$ hold, where $s_i \triangleq \sum_{q=1}^{k_i} w_{i,q} b_{i,q}$.

**Forcing an encumbered history.** This can be supported as in Section 4.2, i.e., by having weights whose sum is close to 1 and V, J fields that define a sum $\sum_{i=1}^{t} w_{i,j_i} b_{i,j_i}$ that is close to the market value of $x$. However, forcing of this kind would require $\texttt{eid} = \texttt{hash}(\text{baseline asset } x \text{ at date } y)$, with $y$ specifying a future date for the target price of $x$.

The formula for assessing the current market price of an asset $z = (m, \texttt{eid}, v, j, \emptyset)$ can be given as
$$\text{price}(z) = m \cdot \frac{1}{k-1} \cdot (1 - \frac{d_j}{s}),$$
where $\texttt{eid} = \texttt{hash}(\text{baseline asset } x)$, $c$ is the current market value of $x$, $v = (b_1, w_1, b_2, w_2, \ldots, b_k, w_k)$, $d_i = w_i \cdot |b_i - c|$, and $s = \sum_{i=1}^{k} d_i$.

Notice that after an initial *outcome-split* of $m$ ordinary coins, it holds that $\sum_{i=1}^{k} \text{price}(z_i) = m$, as it should.

Also note that with this price formula, a fully accurate prediction $b_i = c$ implies earnings of $\frac{m}{k-1}$ ordinary coins.

Using the same denotations, an alternative price formula can be given as

$$\text{price}'(z) = \max(0, m \cdot (1 - (k-1)\frac{d_i}{s}))$$

.

Here $\sum_{i=1}^{k} \text{price}'(z_i) = m$ only when $\forall i : 1 - (k-1)\frac{d_i}{s} \geq 0$, because this formula does not allow the price of an individual share to be negative. This means that someone who holds a share with $1 - (k-1)\frac{d_i}{s} < 0$ has made a very poor bet, but this share is not completely worthless and should be sold on the market for a low price, as it can facilitate an *outcome-combine* transaction.

The upside of the price$'$ formula is that it amplifies the rewards for accurate predictions. In particular, a fully accurate prediction $b_i = c$ results in a maximal earnings of $m$ ordinary coins. This also serves as a demonstration that vector CFDs are capped, as it is impossible to earn more than the initial $m$ coins that were used to create the asset.

In fact, there are infinitely many possible price formulas, since the price is driven by the market, as opposed to being enforced at the protocol level. Thus, it is up to the market participants to pick their preferred price as they see fit, in accordance with the law of supply and demand.

As an example, suppose that Alice transforms 500 coins to $\{z_j = (500, \mathsf{e}_5, v, j, \emptyset)\}_{j=1}^{3}$ with $\mathsf{e}_5 = \mathtt{hash}(\text{baseline asset } x)$ and $v = (75, 1/3, 100, 1/3, 125, 1/3)$. Let us assume that $x$ is currently valued at \$200. Bob predicts that the value of $x$ will fall dramatically, and buys $z_1$ from Alice for $\text{price}(z_1) = 145.8333$ coins. Later, $x$ falls to \$110. Bob sells $z_1$ to Alice for $\text{price}(z_1) = 104.1666$ coins. Alice now executes *outcome-combine* to recover her 500 coins. Hence, Alice collected Bob's loss of $145.8333 - 104.1666 = 41.666$ coins. If $x$ fell further so that its value was closer to \$75 than \$125, Bob would have profited.

Now, in case Carol transforms for example 400 coins to $\{z'_j = (400, \mathsf{e}_5, v', j, \emptyset)\}_{j=1}^{4}$ with $v' = (150, 1/2, 40, 1/4, 50, 1/8, 70, 1/8)$, these shares can take part in the same market with Alice and Bob. For instance, if $z_1$ is divided into $(100, \mathsf{e}_5, v, 1, \emptyset)$ and $(400, \mathsf{e}_5, v, 1, \emptyset)$, then the latter can be combined with $z'_1$ to produce 400 ordinary coins.

## 5 Conclusion

The trust that participants need to extend to different forms of financial services is a spectrum. For a decentralized currency system such as Bitcoin, one can argue that little or no trust is needed. Since the financial instruments that are traded in a prediction market represent only digital information, we motivated and presented a construction for a decentralized prediction market that requires essentially the same level of trust as that of Bitcoin. While our construction readily generalizes to additional financial instruments such as CFDs, other kinds of financial services may require a higher degree of trust.

## A  Real-time semi-decentralized order book

In [7], a fully decentralized order book mechanism is presented. As discussed in [7, Section 6.1], this kind of a decentralized trading platform can work well by letting miners keep the surplus of the spread. However, it is inherently the case that decentralized platforms cannot achieve instant trades when responsiveness to real-time price fluctuations is desired, and that dishonest and self-interested participants can manipulate the market by placing orders and then reneging instead of fulfilling them. Therefore, in the case of a highly liquid PM, a fully decentralized order book might not be the best option for traders.

To complement the construction of [7], we outline an order book mechanism that is semi-decentralized in the sense that traders rely on a supposedly reputable trusted third party (TTP) to execute in real-time the orders that they place, and in case the TTP becomes corrupt they will regain their original assets. That is to say that a corrupt TTP can prevent trades from taking place, but cannot steal the traded assets and disappear.

The basic idea is to let traders deposit assets into a multisignature script that can be spent either by both the trader and the TTP, or by the trader alone but only after a specified time (cf. [23]). Trades are executed off-chain so that the TTP co-signs every transaction and can thus disallow double-spending by malicious traders. Each traded output uses a multisignature script of the above form, so traders are ultimately in control of their assets. From time to time, the TTP publishes the state to the decentralized Bitcoin network, in order to make the off-chain history irreversible.